\documentclass[12pt]{article}
\usepackage{epsfig}
\usepackage{subfigure}
\usepackage{floatflt}
\usepackage{float,hangcaption,xspace}

\newcommand{\registered}{\textsuperscript{\textregistered}}
\begin{document}




\begin{titlepage}
\sf
\pagenumbering{arabic}
\vspace*{-1.5cm}

\vspace{1cm}
\begin{flushright}
{\Large IEKP-KA/04-05}\\
{\Large \today}
\end{flushright}                                                               
\vspace*{1.cm}
\begin{center}
\Large
{\bf \sf
A Neural Bayesian Estimator for \\ 
Conditional Probability Densities  \\
} 
\vspace*{0.3cm}
\normalsize { 
   {\bf \sf Michael Feindt}\\
   {\footnotesize Institut f\"ur Experimentelle Kernphysik}\\
   {\footnotesize  Universit\"at Karlsruhe }\\

   {\footnotesize  michael.feindt@physik.uni-karlsruhe.de }\\
   {\footnotesize  http://www-ekp.physik.uni-karlsruhe.de/
     $\tilde{\mbox{ }}$feindt} \\
   
}
\end{center}
\vspace{\fill}
\begin{abstract}
\noindent

This article describes a robust algorithm to estimate 
a conditional probability density $f(t|\vec{x})$ as a non-parametric
smooth regression function. It is based on a neural network and the
Bayesian interpretation of the network output as a posteriori
probabability. The network is 
trained using example events from history or simulation, which 
define the underlying probability density $f(t,\vec{x})$. 

Once trained, the network is applied on new, unknown examples 
$\vec{x}$, for which it can predict the probability distribution 
of the target variable $t$. Event-by-event knowledge of the smooth
function $f(t|\vec{x})$ can be very useful, e.g. 
in maximum likelihood fits or for forecasting tasks. No assumptions
are necessary about the distribution, and non-Gaussian tails are
accounted for automatically.
Important quantities like median, mean value, left and 
right standard deviations, moments and
expectation values of any function of $t$ are readily
derived from it.

The algorithm can be considered as an event-by-event unfolding and leads
to statistically optimal reconstruction.
The largest benefit of the method lies in complicated problems,
when the measurements $\vec{x}$ are only relatively weakly correlated to 
the output $t$. As to assure optimal generalisation features and to avoid
overfitting, the networks are regularised by extended versions of
weight decay. The regularisation parameters are determined during the
online-learning of the network by relations obtained from Bayesian
statistics. 

Some toy Monte Carlo tests and first real application examples from 
high-energy physics and econometry are discussed.


\end{abstract}
\vspace{\fill}
\begin{center}

This article is a reprint of an internal EKP note from January 2001,
corrected and supplemented by information on the development in the
3 years since then. The algorithm is implemented and further developed
in the NeuroBayes\registered package by Phi-T\registered 
Physics Information Technologies GmbH, Karlsruhe, Germany.

\end{center}
\vspace{\fill}
\end{titlepage}
\sf





\section{\sf Introduction}
Assume we have a random variable $t$ whose probability density 
function is known
or can be estimated from a large but finite number of examples, e.g. the 
energy distribution of a certain very short-lived
particle in a high energy physics experiment or the daily change of 
an equity price. The aim is to model the probability density 
{\sl for a given event or date} from available input data. 
In our examples one might have one or more,
perhaps correlated, not very precise measurements of the energy. Or one 
knows the actual price and the recent history plus some technical indicators
at a given day and wants to predict what will be the probability density
for next day's or week's price change. It is easy to know what
happens on average, given by $f(t)$, but one wants to have a better 
estimate taking into account the particular 
situation of the event under consideration, $\vec{x}$.

We do not just want to have a single number for $t$, but an estimate 
of the complete probability density, from which we then can deduce 
the most probable value, the mean value, the median, moments, but also
uncertainty intervals or expectation values of any function of $t$. 
As an example, from the probability density describing an 
equity price change one can compute the probability density of an option
price for that equity and its fair price.

Precise measurements are easy to handle, one just 
identifies the measured value with the true value and uses classical
statistical methods for error estimates etc.
The method proposed in this article has its largest benefit for difficult 
predictions, 
when the measurements are not very strongly correlated to the desired output,
i.e. when the individual measurements are not assumed to be
much more precise than the width of the complete distribution
averaged over many events. 
This certainly is the case in the last example of predicting future 
equity prices.
  
The algorithm is a Bayesian estimator in the sense that it takes into 
account a priori knowledge in the form of the inclusive (unconditional) 
distribution.
It will never result in unphysical values outside the training range, a
very helpful feature when the input measurements are not very exact. 
The influence of the shape of the inclusive distribution diminishes with
increasing precision of the input measurements \cite{sivia}.
Bayes theorem plays a large role in two other places: the interpretation 
of network output levels as Bayesian a posteriori probabilities in 
classificiation problems, and for determining regularisation constants.

In recent years the analysis methods in High Energy Physics 
experiments have been steadily further developed 
\cite{blobel, cowan}. Neural network algorithms for classification 
tasks \cite{nnliteratur} have become very successful and - 
after initial scepticism in the community - essentially accepted. 
In the DELPHI experiment at LEP the author and coworkers have optimised 
electron identification 
\cite{ELEPHANT}, kaon and proton identification \cite{MACRIB} using 
neural networks, and developed inclusive b-hadron reconstruction 
algorithms \cite{BSAURUS} for energy, decay length, lifetime reconstruction, 
B hadron identification, particle/antiparticle distinction at production 
and decay time etc., which have been and are applied in many analyses. 
These are hybrid algorithms combining classical approaches with a large 
number of neural networks at different levels (single track, hemisphere, 
event). 
Also other modern statistical methods have been investigated, e.g. 
Support Vector Machines \cite{vapnik}. However, we have found 
\cite{supvec} that for difficult non-separable classification problems often 
encountered in High Energy Physics simple neural networks with intelligent
preprocessing are more useful. With this knowledge and experience we have
searched for a method to extend this technology to continuous, real-valued 
variables $t$.

The proposed algorithm is based on a simple feed-forward neural network 
which is trained using backpropagation 
\cite{backpropagation}
of either simulated
Monte Carlo or historical data. The cumulated probability distribution
of the whole dataset is discretised into $N$ bins, with variable
width but same amount of statistics inside. For each of the bins, a separate 
output node is trained to the binary classification problem 
``the true value is above the threshold value'' vs. ``the true value is below
the threshold value''. The output of the network is
filtered through a symmetric sigmoid transfer function, as usual in binary 
decision nets.
%
%
%
%
A cubic B-spline (see e.g. \cite{blobel}) is fit through the $N$ filtered net 
output values, forced 
through $-1$ and $1$ at the extreme values and applying Tikhonov-type
regularisation \cite{tikhonov} on the basis of the third derivatives' squares.  

This spline, properly rescaled, is a robust estimator of the 
cumulative probability distribution function of the true value for 
a given event. 
Median and quantiles are easily obtained from it, 
the derivative of the spline is an estimator of the 
probability density itself. This can e.g. be employed in maximum 
likelihood fits. Reconstructing the truth distribution for each individual
event, the algorithm corresponds to an event-by-event unfolding. 
In fact, it can be extended to an unfolding procedure to estimate an
unknown $f(t)$, but this is left for future research.

Some methods to predict conditional probability distributions from 
examples can already be found in the literature, but have been come to the
author's attention only after this work was essentially completed. 
Weigend's and Srivastava's ansatz \cite{weigend} is similar to the 
one presented
in this paper, also based on a feed-forward network with several output nodes,
however it works directly on the probability density function. It does not
have the aim of a smooth density function, does not include regularisation 
terms and does not include pre- and postprocessing.
Other methods are based on kernel estimation \cite{kulczycki,vapnik}.

Section 2 contains a detailed description of the algorithm, in
section 3 the theoretical background is laid out. 
In section 4 we review and develop a number of optimisations 
implemented during the development of the example nets.
Section 5 and 6 describe toy and real world example applications from 
high energy physics and econometry.     

\section{\sf The algorithm}

Assume we have a random variable $t$ which is distributed according 
to a probability density $f(t)$. $t$ may e.g. be the day-to-day change of an 
equity price or a quantum physical quantity.
Our knowledge of this density is empirical, i.e. it is determined by 
a large number of examples, e.g. historical data or data simulated 
by Monte Carlo methods.
For each of these examples, or ``events'', labelled $i$,  there 
exists a vector of measurements $\vec{x_i}$ that are correlated to 
the variable $t_i$. An overall probability density $f(t,\vec{x}$) is
defined by the training sample.

It is the aim of the method described in this paper to achieve a smooth
estimate of the conditional probability density $f(t|\vec{x_i})$ for a 
given measurement vector $\vec{x_i}$. If there is no information in the 
input vectors then obviously $f(t_i|\vec{x_i})=f(t_i)=f(t)$, i.e. the
conditional probability is equal to the inclusive distribution.
This would e.g. be true for a random walk model in econometry. 
If however there is a correlation then one should be able to get a better 
estimator for a given event, $f(t_i|\vec{x_i})$, and from that one can 
deduce better expectation values and error intervals.

The method proceeds in several steps:

\subsection{\sf Preprocessing of the target values} 
\begin{figure}[!th]
\mbox{\epsfxsize16.0cm\epsffile{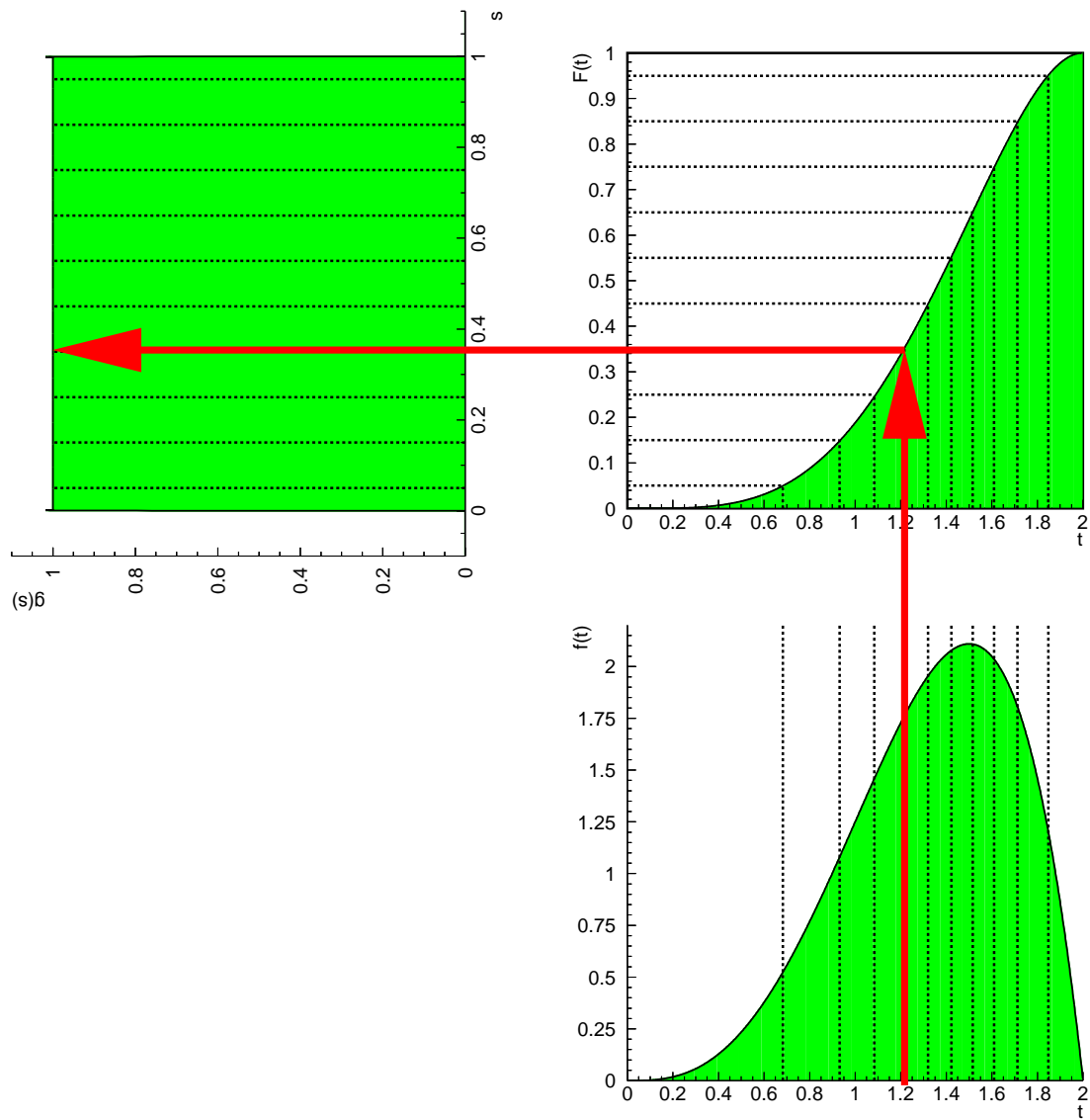}}
\caption[]
{Variable transformation $F:t\to s $ (upper right) 
leading to a flat 
output distribution $g(s)$ (upper left) when $t$ is distributed according
to $f(t)$ (lower plot).}
\label{transformation}
\end{figure}

 At first we perform a flattening of the distribution $f(t)$ by an 
 in general non-linear, monotonous variable transformation $F:t\to s$, 
 such that the transformed variable $s$ is distributed uniformly between 
 0 and 1: $s(t_{min})=0$ and $s(t_{max})=1$:
 \begin{equation} 
s = F(t)= \int_{t_{min}}^t f(t') dt'.
 \end{equation}
 $s$ measures the fraction of events that has lower values of $t$ than
 the actual $t$. 
 The probability density in the transformed variable $g(s)$ is simply 
 constant $1$ in the interval $[0,1]$. The value $s$ therefore can be 
 interpreted as the cumulative probability density of its own distribution:
\begin{equation} 
s= G(s) = \int_0^s g(s') ds'.
\end{equation}
This is illustrated in Fig.~\ref{transformation}.

A robust method of constructing $F$ from a finite number of examples
is to sort them (using a binary search tree) 
according to $t$. $s(t_i)$ then just is $i/N_{events}$, 
when $t_i$ is the $i^{th}$ element in the sorted list.
It is practical to store the $t$-values in $1\%$ bins, i.e. the 
$101$ values $t_{min},t_{1\%},t_{2\%},...,t_{max}$. 
The functions $F$ and $F^{-1}$ for transformation and back transformation are 
constructed using this list and linear interpolation.
However, for some problems there may be long but important tails
(e.g. describing stock market crashes) such that by just storing $1\%$
quantiles information may be lost. To avoid that, a histogram 
with $200$ equally sized bins between the observed minimum and maximum 
value is stored in addition. Later both lists are used for 
reconstruction.  

\subsection{\sf Discretisation}  
The probability distribution $G(s)$ is now sampled at $N$ equidistant
levels $L_j=  (j-0.5)/N $, $j=1,N$.
From the property $G(s)=s$, it further follows that $G(0)=0$ and
$G(1)=1$, such that $N+1$ intervals are defined. All the inner intervals
contain the same number of events, namely $N_{events}/N$, the first and
last interval half of this number. For $N=10$ this is illustrated in Fig. 1. 

\subsection{\sf Neural Net Prediction}
\begin{figure}[!th]
\mbox{\epsfxsize16.0cm\epsffile{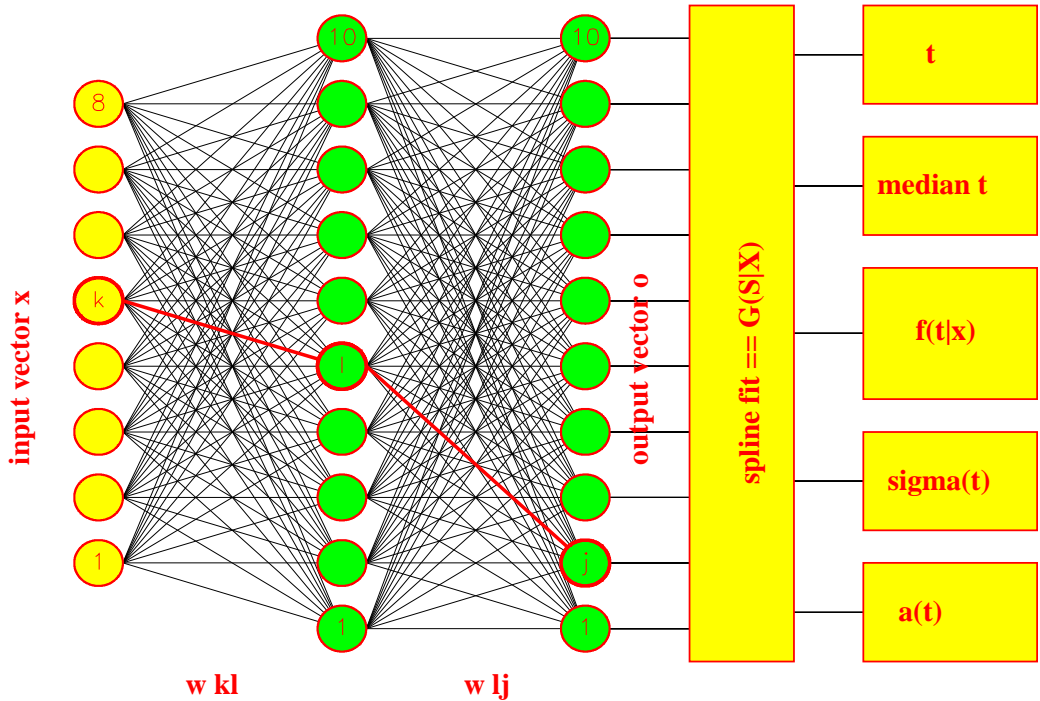}}
\caption[]
{Example architecture of the Bayesian neural net, with 8 input nodes,
10 intermediate, and $N=10$ output nodes. The lines denote the
connections, each of which is associated with a weight that is
optimised in the training. 
The post-processing consists of a regularised spline fit through the 
single nodes' output, from which the interesting final quantities can 
be calculated.  
}
\label{architecture}
\end{figure}

Classification of events into two classes is a task that neural networks 
can handle well. A possible classification task for a neural network is
to separate events with $t$ larger than a threshold value $L$ from those
with $t<L$.
The main idea is that the conditional cumulated probability density 
$G(s|\vec{x})$ can be estimated from many neural networks that perform
such classifications at different thresholds $L_j$. Instead of training
several independent networks, this is done in one net that has 
$N$ output nodes, corresponding to the $N$ discretised output levels. 
In such a (usually sufficient) three layer neural net (see Fig. 2) the 
$N$ outputs $o_j$ are calculated via

\begin{equation} 
o_j = S(\sum_l w^{2\to 3}_{lj} \cdot S( \sum_k {w^{1\to
      2}_{kl} \cdot x_k } )  ) 
\end{equation}

where the $x_k$ are the input values and the index $l$ runs over the 
intermediate layer nodes.  $w^{1\to 2}_{kl}$ is the weight
of the connection between node $k$ of the first and node $l$ of the 
second layer, and  $w^{2\to3}_{lj}$ that between node $l$ of the 
second and node $j$ of the output layer. $S(x)$ is a transfer function 
that maps $]-\infty,\infty[$ to the interval $[-1,1]$, for which we apply 
the symmetric sigmoid function  
\begin{equation} 
S(x) =  \frac{2}{1+e^{-x}} -1.
\end{equation}
The architecture of the network is shown in Fig.\ref{architecture}.

As in common simple classification neural net training, 
the weights are determined using the iterative back propagation 
algorithm employing historical or simulated data, with some
acceleration and regularisation techniques applied as outlined 
in section 4. 

Each output node $j$ is trained on the binary decision: is the real 
output $s_{true}$ larger than $L_j$ (target value = $1$) or it is smaller
(target = $-1$)?
Thus the target vector of an event with $s_{true}=0.63$ and $N=10$ levels
would be $ \vec{T} = ( +1, +1, +1, +1, +1, +1, -1, -1, -1, -1,)$.  
Note that $s_{true}=0.56$ and $s_{true}=0.64$ would give the same target
vector, this is the discretisation uncertainty introduced. If the individual
measurement resolution is good, this is a potential source of information
loss. The number of intervals $N$ should be matched to the obtainable 
resolution.
An alternative to reduce the discretisation information loss in the 
training is to set the target node nearest to the true value $s_{true}$ to 
a value between $-1$ and $1$ as to smoothen the dependence. In this 
training mode the target vector would be 
$ \vec{T} = ( +1, +1, +1, +1, +1, +0.2,  -1, -1, -1, -1,)$ for 
$s_{true}=0.56$, but 
$ \vec{T} = ( +1, +1, +1, +1, +1,   +1,-0.2, -1, -1, -1,)$ 
for  $s_{true}=0.64$. 
This procedure can be used when employing the quadratic loss function.

\begin{figure}[!th]
\mbox{\epsfxsize16.0cm\epsffile{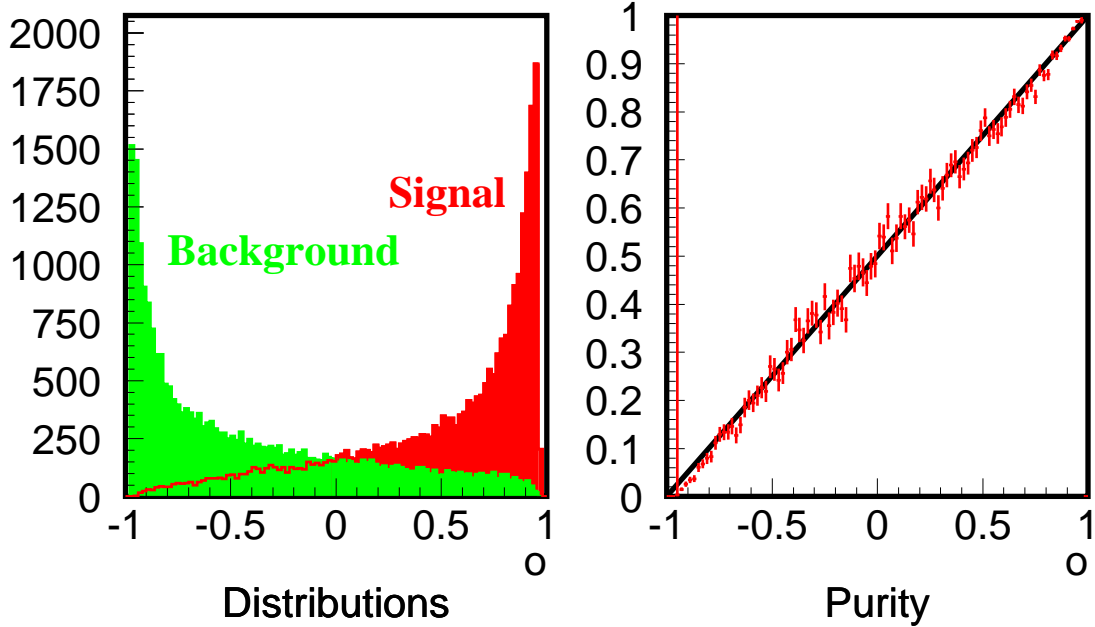}}
\caption[]
{Left: Signal (true $s$ is larger than $L_{10}=0.475$) and background 
distribution of
the $10^{th}$ node of a $N=20$ network. Right: 
Signal purity in each bin as function of network output $o$. 
The expected linear behaviour is observed, demonstrating that 
the net is well trained.}  
\label{purity}
\end{figure}

\subsection{\sf Neural Net Output Interpretation}
After minimising the quadratic loss function 
\begin{equation}
\chi^2 = \sum_j w_j \chi_j^2 = \sum_j w_j \frac{1}{2}\sum_i (T_{ji}-o_{ji})^2,
\end{equation}
(where $T_{ji}$ denotes the target value for output node $j$ in event $i$  
and the definiton and role of the $w_j$ is as described in section 3)
in the network training, and assuming that the network is well trained, 
the network output $o_j$, rescaled to the interval $[0,1]$, 
is equal to the purity 
\begin{equation}
P(o_j)= f(s_{true}<L_j|o_j)/f(o_j)  = (o_j+1)/2.    
\end{equation}
This can be seen as follows: The mean $\chi^2$ contribution of measured 
events with output $o$ is 
\begin{equation}
\chi^2 =  P\cdot (1-o)^2  + (1-P) \cdot ((-1)-o)^2      
\end{equation}
where the first term describes the contribution of signal events 
(purity $P$, target $+1$), and the second of background events
(purity $1-P$, target $-1$).
When the network is trained optimally,
$\chi^2$ is minimal, i.e. $d\chi^2 / do = 0$.
This directly leads to $P=(o+1)/2$. 
Figure \ref{purity} shows signal and background distribution and the purity
in each bin 
as a function of the net output for the $10^{th}$ node of a $N=20$ network.
Thus $ (o_j+1)/2 $ is the probability that the true value 
really lies below the given threshold $L_j$.
An interpolation through the $N$ rescaled network outputs is the
desired conditional cumulated probability density $G(s|\vec{x})$.

The same arguments also hold for the entropy loss function:
\begin{equation}
\label{entropyloss}
E_D = \sum_j w_j E_D^j = \sum_j w_j \sum_i 
\log( \frac{1}{2}(1+T_{ji}\cdot o_{ji}+\varepsilon))
\end{equation}
which we prefer since it has some advantages in classification problems.
In this case a completely wrong classification $o_j=1$ for $t_j=-1$ or
vice versa leads to an infinitely large $E_D$. To get rid of such completely
wrong classifications is thus the first thing learned by the network
using the entropy error function. In order to avoid numerical problems
for untrained networks, a small regularistaion constant $\varepsilon$
is introduced. $\varepsilon$ is reduced in each training iteration and
is zero after just a few iterations.    

Moreover, the absolute value of $E_D$ has a meaning in Bayesian statistics 
avoiding an extra regularisation constant in the weight decay 
regularisation scheme (see section~\ref{regularisation}).

\subsection{\sf Interpolating the discrete network outputs}
In order to get a closed functional form and to be able to calculate
$G(s,\vec{x})$ and its derivative, the probability density $g(s,\vec{x})$, 
at any value $s$, a cubic B-spline fit with variable bin size is 
applied through the
end points $(0,1)$, $(1,-1)$ and the
$N$ points $(L_j,o_j)$ estimated by the neural network. 
It is constructed
from a 4-fold knot at $s=0$, a 4-fold knot at $s=1$, and  a number of
simple knots placed in between $0$ and $1$. The actual number should be chosen
smaller than the number of output nodes $N$, as to smoothen statistical 
network fluctuations. It has been proven useful to not distribute the nodes
equidistantly, but to place more in the regions of a large third derivative.

We want to achieve a smooth distribution function $g(s|\vec{x})$. This
is achieved by the following Tikhonov type regularisation procedure:
The total curvature $ c= \int_0^1 (g(s|\vec{x})'')^2 ds $ should be as
small as possible with $G(s|\vec{x})$ still giving a good description of the
$N$ output levels. Constructing $G(s|\vec{x})$ using cubic B-splines, $c$ is
just the bin-size weighted sum of the third derivatives' squares of the spline.
These are constants between any two knots of the spline, and can be easily
added to the normal equations.  

Thus the regularised spline fits stay linear fits which only demand an
inversion of a symmetric $N \times N $ band matrix of band width $4$, which
is performed very quickly \cite{blobel}. 
In addition, in order that $F(t|\vec{x})$ can be interpreted as 
cumulated probability distribution, the constraints of 
monotoniticity, positivity and a maximum value of 1 must be satisfied.
In order to keep the fits linear and fast, these unequality constraints 
are fulfilled by the following procedure:
First a normal fit is performed , this can be done analytically in
one step. If the result does not fulfill all requirements,
a quadratic loss function multiplied by a large number is added to 
the normal equations of the 
corresponding parameter or difference of parameters. This way a new fit 
is expected to result at exactly the border of the allowed region. 
Then the procedure is iterated, until all constraints are fulfilled. 
Accepting numerical uncertainties of $10^{-6}$, one or two iterations 
are usually enough. This procedure works well and is considerably
easier to program than iterative usage of Lagrange multipliers to
fulfill the unequality constraints.    

The probability density 
$g(s|\vec{x})$ can be calculated analytically from the spline
parametrisation.
To transform back to the original variable $t$ the inverse mapping $F^{-1}$
has to be applied. If $F$ is stored as $101$ interval borders as described 
above, $t$ can be retrieved from searching in the list and linear 
interpolation.
The function $f(t|\vec{x})=f(t)\cdot g(F(t),\vec{x})$ also is calculated
as a 100 value table with linear interpolation.
  
\subsection{\sf Important Quantities:}
The {\bf median} of the conditional distribution $f(t|\vec{x})$ is calculated 
from 
\begin{equation}
t_{med}=F^{-1}(G^{-1}(0.5)).
\end{equation}
{\bf Left and right error intervals} may be defined as in Gaussian 
distributions
as the limits of the interval that contains $68.26\%$ of the data: 
\begin{equation}
\sigma_{left}=F^{-1}(G^{-1}(0.5))-F^{-1}(G^{-1}(0.8413))
\end{equation}
and
\begin{equation}
\sigma_{right}=F^{-1}(G^{-1}(0.1587))-F^{-1}(G^{-1}(0.5))
\end{equation}
The expectation or {\bf mean value} $\langle t \rangle$ can be estimated from
\begin{eqnarray} 
\langle t \rangle & = & \int t' f(t'|\vec{x}) dt' \\
                  & = & \int t'(s') g(s'|\vec{x}) ds' \\
             & \approx& \sum_j F^{-1}(s_j) g(s_j|\vec{x}) \\ 
             & \approx& 1/M \sum_{m=1}^{M} F^{-1}((G^{-1}((m-0.5)/M))
\end{eqnarray}
This latter expression is particularly simple to calculate:
One just chooses M equidistant points $y_m$ between 
$0$ and $1$, and performs the two back transformations $F^{-1}(G^{-1}(y))$
successively on them to achieve the corresponding $t_m$ values, 
and takes their average.
This simple operation even is an optimal estimator of the integral
at a given number of function evaluations: It contains an implicit 
importance sampling. 

The {\bf expectation value for
any function $a(t)$ } of $t$ is estimated from 
\begin{eqnarray} 
\langle a(t) \rangle & = & \int a(t') f(t'|\vec{x}) dt' \\
         &\approx &\sum_j a(F^{-1}(s_j)) g(s_j|\vec{x}) \\ 
 & \approx& 1/M \sum_{m=1}^{M} a(F^{-1}((G^{-1}((m-0.5)/M)))
\end{eqnarray}
Again the last expression delivers a very simple and effective way of
calculation. An example application is option price calculation, leading
to a better prediction than the Nobel-prized 
Black-Scholes formula \cite{blackscholes}.

\section{\sf Theoretical framework}
Here we shortly summarise the mathematical framework 
from statistics and learning theory of the heuristic method
explained in the previous section.

\subsection{\sf The learning problem} 
A probability density $f(t)$ is defined from the following 
Fredholm integral equation
of first kind: 
\begin{equation}
\int_{-\infty}^t f(t') dt' = F(t).
\end{equation}
The probability distribution function $F(t)$ for a scalar random variable is
defined as the probability that a random realisation has a value smaller than
$t$: 
\begin{equation}
F(t)=P\{\xi<t\}.   
\end{equation}
For a vector quantity, $F(\vec{x})$ is defined coordinatewise:
\begin{equation}
F(\vec{x}) =
\int_{-\infty}^{\vec{x}} f(\vec{x'}) d\vec{x'} 
       = \int_{-\infty}^{x_1} ...\int_{-\infty}^{x_n} 
        f(x_1',...,x_n') dx_1' ... dx_n'. 
\end{equation}
A {\em conditional} probability density $f(t|\vec{x})$ is defined by
\begin{equation}
\label{cp}
\int_{-\infty}^{t}\int_{-\infty}^{\vec{x}} f(t'|\vec{x'}) dF(\vec{x'}) dt' = F(t,\vec{x}).
\end{equation}
Here $F(\vec{x})$ is the distribution of random vectors $\vec{x}$, and 
$F(t,\vec{x})$ is the joint distribution function of pairs $(t,\vec{x})$.
Since $dF(\vec{x})=f(\vec{x}) d\vec{x}$ and 
$f(t|\vec{x}) = f(t,\vec{x})/f(\vec{x})$, 
$f(t|\vec{x})$ is a conditional probability
density. Equation \ref{cp} is an example of a linear operator
equation  $ A f = F $, where both the right side $F$ and the operator
$A$ (depending on $F(\vec{x})\approx F_l(\vec{x})$) are only 
approximately known.

Our problem is to estimate a non-parametric solution $f(t|\vec{x})$
of integral equation \ref{cp}, where the probability
distributions $F(\vec{x})$ and  $F(t,\vec{x})$ are unknown but defined
from a finite number of pairs
$(t_1,\vec{x_1}),...,(t_l,\vec{x_l})$.
This means that $F(\vec{x})$ and $F(t,\vec{x})$ are approximated
by the empirical distribution functions
\begin{equation}
F_l= \frac{1}{l} \sum_{i=1}^l \Theta(t-t_i) \cdot \Theta(\vec{x}-\vec{x_i})
\end{equation}
where 
\begin{equation}
\Theta(x) = \left\{ \begin{array}{cl} 0 & \mbox{for}\,\,\, x < 0 \\ 1 & \mbox{for}\,\,\, x\ge
0\end{array} \right. 
\end{equation} 
denotes the Heaviside step function.
The Glivenko-Cantelli theorem \cite{glivenko} assures that this empirical
distribution function $F_l$ is a good estimate of the actual distribution 
function $F$, and uniform convergence takes place as $l\to \infty$.
Kolmogorov and Smirnov have shown that 
the convergence is proportional to $\sqrt{l}$
\cite{kolmogorov}.
For one-dimensional probability distributions, the deviation is limited by 
\begin{equation}
 \lim_{l\to\infty} P\{ \sqrt{l} \sup_t | F(t)-F_l(t) | \ge \varepsilon \} 
= e^{-2\varepsilon^2}. 
\end{equation} 
For the multidimensional case the limit is unknown but
known to exist \cite{vapnik}.

\subsection{\sf Solving the ill-posed problem}

To solve the integral equation the quadratic risk function 
$W = || Af - F ||^2$ is minimised. 
Since we approximate the real risk by the risk calculated from the given 
events, one talks about empirical risk minimisation (ERM).
 
Solving such a Fredholm integral equation of first kind 
is known to be an ill-posed problem. 
This means that it cannot be assured
that the solution is stable. An infinitesimal change on the right hand 
side may 
lead to a large variation in the solution $f(t|\vec{x})$. Oscillatory 
behaviour is a typical phenomenon of solutions of ill-posed problems.
Several regularisation schemes have been proposed to stabilise the solutions,
putting in some theoretical prejudice. 
One of the most common techniques is the Tikhonov regularisation scheme
\cite{tikhonov}.
Here the risk function to be minimised is modified to 
$ W_T = || Af-F ||^2 + \gamma \Omega(f)$ 
by adding a regularisation term.
A possible  $\Omega(f)$ could be the total curvature of the solution $f$
if one knows it to be a smooth function. 
The regularisation parameter $\gamma>0$ has to be chosen such that 
the original risk $W$ is still small and simultaneously good
smoothness is obtained.

Another important ansatz is structural risk minimisation (SRM) \cite{vapnik}.
Even in so-called {\em non-parametric} estimations the solution is
constructed from a 
number of basis functions, e.g. polynomials, splines, or some orthogonal 
function set. Learning then means to determine the free
parameters that e.g. describe the amount of each basis function.

\subsection{\sf Generalisation error and VC dimension}
Learning theory has shown that generalisation ability is best when only
few parameters are necessary. However, not only the number of parameters
is decisive, but also the structure of the functions. So high frequency
oscillatory functions may contain only few parameters but can lead to very
unstable solutions. Vapnik and Chernovenkis \cite{vc} have introduced 
the concept of the VC dimension. A small VC dimension means good 
generalisation ability of a learning machine, whereas infinite VC dimension 
denotes just a learning by heart of the given events and no generalisation 
ability at all.
For affine linear functions in $n$-dimensional space, the VC dimension simply
is $n+1$. This is not true in general, for non-linear functions  
it may be much larger.

For neural network training, a structure may be defined by the network
architecture.
The more hidden nodes are introduced, the smaller the empirical risk, but
the larger the structural risk. Overfitting may occur and generalisation 
ability degrades when too many nodes are used. This requires
careful control. 

The VC dimension of neural networks can decrease significantly when 
the weights are bounded to be small \cite{boser}.
However, independent of the size of the weights, a lower limit on the 
VC dimension of a three-layer neural network 
with $n$ input nodes and $k$ hidden nodes has 
been established \cite{leebartlett}:
\begin{equation}
VCdim \ge (k-1)\cdot(n+1),
\end{equation}
equal to the number of weights of a network with $n-1$ input nodes  
and $k-1$ hidden nodes. VC dimension is a worst case concept, and 
there is substantial evidence that  
improved generalisation performance is achieved when networks are trained
with weight decay regularisation \cite{krogh}. This can be understood
in a Bayesian framework \cite{mackay}, which also allows to choose 
regularisation parameters $\alpha_c$ and $\beta$:
\begin{equation}
M = \sum_c \alpha_c E^c_W(\vec{w}|{\cal A}) + \beta E_D(D|\vec{w},{\cal A})
\end{equation}   
where
\begin{equation}
\label{quadraticloss}
E_D(D|\vec{w},{\cal A}) = 
\sum_{events\,m} \sum_{output\,nodes\,i}\frac{1}{2}
\left( o_i(\vec{x}^m)-t_i^m \right) ^2
\end{equation}
is the quadratic error function of the data and
\begin{equation}
E^c_W(\vec{w}|{\cal A}) = \sum_{i\,weights} \frac{1}{2} w_i^2
\end{equation}   
is the regularisation energy (which leads to a subtraction of a multiple of
$\vec{w}$ in the weight update rule). It has been shown\cite{mackay}
to be useful to separate
this term into three ($c=1,2,3$) terms, which determine the weight decay of
the weights of the input measurements to the second layer, the bias node to the
second layer, and the second to the output layer, since there is no
a priori reason why these weights should be in the same order of magnitude.   

\subsection{\sf Bayesian approach to regularisation} 
\label{regularisation}
The Bayesian approach in \cite{mackay} leads to the following optimal 
parameters:
The point of maximal probability $P(D|\alpha,\beta)$ to observe 
the observed data $D$ has the following property:
\begin{eqnarray}
\chi^2_W\equiv 2\alpha E_W &=& \gamma \\
\chi^2_D\equiv 2\beta  E_D &=& N-\gamma 
\end{eqnarray} 
where $\gamma$ is the effective number of degrees of freedom: 
\begin{equation}
\gamma = \sum_{a=1}^k\frac{\lambda_a}{\lambda_a+\alpha}
\end{equation}
where $\lambda_a$ are the eigenvalues of the quadratic form $\beta E_D$ in the
natural basis of $E_W$. 
If the entropy function eq.~\ref{entropyloss} is used instead of the quadratic 
loss function (\ref{quadraticloss}), there is no need for a $\beta$ to be
estimated from the data \cite{mackay2}.    

Calculating the Hessian matrix 
$H=\nabla \nabla M$
allows to distinguish between well and poorly determined parameters, 
and to calculate error bars on weights and net outputs. This is not yet 
implemented in NeuroBayes.

\subsection{\sf Pruning}
When during the training weights become completely insignificant
(less than $0.001\sigma$), the connections are pruned away, i.e. set
to exactly zero. Thus, the architecture is changed and the number of
free parameters is lowered. The VC-dimension explicitely is reduced by
this procedure.
It is interesting to note that something
similar works in biology: neurobiologists have found that the
intelligent mature brain
has less connections between the neurons. In contrast,
an untrained (young) neural network still has many connections.
During its development the brain actually loses the
neuronal connections that are less used, and forms strong connections
in those synaptic circuits that have been utilized the most.  Pruning
improves the signal-to-noise ratio in the brain by removing the cause
of the noise. This process is constant and quick. Synaptic
connections can form in a matter of hours or days.  Experience,
particularly in childhood up to early adulthood, sculpts the brain
\cite{pruning}.

\subsection{\sf Automatic Relevance Determination}
MacKay \cite{mackay} has introduced an interesting extension of the 
weight decay procedure by introducing a hyperparameter for each 
individual input node. This is called ``automatic relevance 
determination'' and allows the algorithm to better prune away unsignificant 
input variables. We have found this extension to be useful.
 
\subsection{\sf Automatic Shape Smoothing}
A direct copy of the same concept, 
for the connections to the different output
nodes, called Automatic Shape Smoothing, 
also is implemented in NeuroBayes.

\section{\sf Some considerations for efficient net training}
Here we list a few experiences we have gained during development of this
code.

\subsection{\sf Online instead of batch training mode}
The training is performed in stochastic, mini-batch or quasi-online
mode: A weight update is performed after about 200 events.
Since the problems to be solved by NeuroBayes typically learn from
relatively large and partly redundant training sets, online learning
is much faster than batch mode, where weight updates are only performed
after all training examples have been read \cite{efficient}. 

\subsection{\sf Random start values for weights}
We preset all weights with random numbers distributed around
zero. To ensure a fast initial learning, it is useful to take Gaussian
random numbers with a $\sigma=1/\sqrt{n_{in}}$, where $n_{in}$ is the
fan-in of a neuron, i.e. the number of incoming weights.  
If the input variables are also distributed like a standard Gaussian
(see below), then this is true also for the output of the hidden layer 
nodes, such that the same argument holds for the output layer. 
This ensures optimal learning conditions directly from the start.

\subsection{\sf Non-randomness of start values in the second and output layer}
However, since we expect in each single event that the net output $o_j$ is a
monotonous function of $j$, we start off with the same random weight
$w^{2\to 3}_{lj}$ for all $j$ for a given $l$. During the training they will
quickly change, but due to the monotonous targets of all training 
events this feature will survive at each training step and thus reduce
statistical uncertainties in the reconstruction of the probability 
density function.

\subsection{\sf Relative importance of the N output nodes}
What are the maximum errors that the single output nodes can contribute?
Using the quadratic loss function,
this can be estimated from 
\begin{equation}
\chi_j^2 \approx N \int_{-1}^{1} g(o_j) 
( P_j\cdot (o_j-1)^2 + (1-P_j)\cdot (o_j+1)^2 ) 
\end{equation}   
where $g(o_j)$ is the distribution function of output node $j$ and
the purity $P_j$ is the fraction of events with target $+1$.
In an untrained net with random weights, when the outputs still are 
completely random,
$g(o_j)=1/2$, the mean is $\chi^2/N=4/3$ independent of $P_j$.

The next extreme is that the network learns the $target =+1$ to 
$target=-1$ ratio for each output node, but nothing else. This corresponds
to $g(o_j)=\delta(o_j-2P_j+1)$ and a mean $\chi^2/N=4P_j(1-P_j)$, which
is $1$ for $P_j=50\%$, and down to $0.0975$ at $P_j=2.5\%$ and $97.5\%$.  
  
This is a trivial learning, i.e. the net learns the inclusive distribution,
but no improvement from the individual input vectors. It thus is useful
to know about this (constant) contribution from $\chi_{j,inclusive}^2$. 
To follow and judge the quality of the training we compare each node's 
$\chi_j^2$ to the $\chi_{j,inclusive}^2$. 
Ratios below $1$ indicate real learning beyond the shape
of the inclusive distribution. Sometimes
it has been useful to give a larger weight to the extremal nodes
after some iterations during the backpropagation step, such that small 
systematic effects here have a 
chance to be learned compared to the large statistical fluctuations 
of the inner nodes.  

This is not necessary with the entropy error function. In this case, 
learning
just the inclusive distribution corresponds to 
$E_D/N=  P_j\log{(P_j)} + (1-P_j)\log{(1-P_j)}$.

\subsection{\sf Learning with weighted events:}
It is possible to train the network with weighted events.
This is useful if e.g. one set of Monte Carlo simulation events is available 
but single input distributions (e.g. lifetimes or branching ratios) must
be adjusted to new (better) values. For time series prediction one may want
to give more recent data more weight than ancient data, without completely
neglecting the latter.

A too large variation of the weights however drastically reduces the
statistical accuracy of the learning, the ``effective'' number of training
events decreases, and statistical fluctuations of the large weight events
may start to fool the network.

The backpropagation algorithm is adjusted by multiplying the 
difference of output and target by the weight. The mean weight should be
one.

\subsection{\sf Bias nodes}
The introduction of bias nodes, i.e. nodes which have a constant value
$1$, are generally extremely useful for net learning. A bias node in the
input layer is clearly advantageous. 
However, in this multi-output net the existence of a bias node in the 
second layer has proven not to be useful and may lead to convergence into
a non-optimal local minimum. The reason is that at the beginning
of the training the error is dominated by the 
vastly different number of signal and background events for the non-central
output nodes. For example, just $2.5\%$ 
of the data has truth values below the first threshold for $N=20$ levels.
The fastest way to learn this fact is to shift the threshold of the final 
sigmoid transfer function to $-1.9$, which is easily achieved when a bias
node is available in the second layer. Only later the other variables are 
looked at, however now at the expense that often the relevant output node 
already is saturated. Thus, learning is more difficult. In practice, 
leaving out a bias node in the intermediate layers always has led to
smaller $\chi^2$ values.
\subsection{\sf Shifts in individual output transfer functions}
We noticed that at least in difficult (low correlation) cases it may
be of advantage to shift the weighted sum of the output units such that,
if the weighted sum is zero, the inclusive distribution results automatically.
This is achieved by the following replacement of the argument of the
transfer function of the output node $j$: 
\begin{equation}
S(\sum{w_{ij}a_i}) \to  S(\sum{w_{ij}a_i} +S^{-1}(1-2*P_j)) 
\end{equation} 
where $P_j$ denotes the nominal probability level of the inclusive 
distribution.  

\begin{figure}[!th]
\mbox{\epsfxsize16.0cm\epsffile{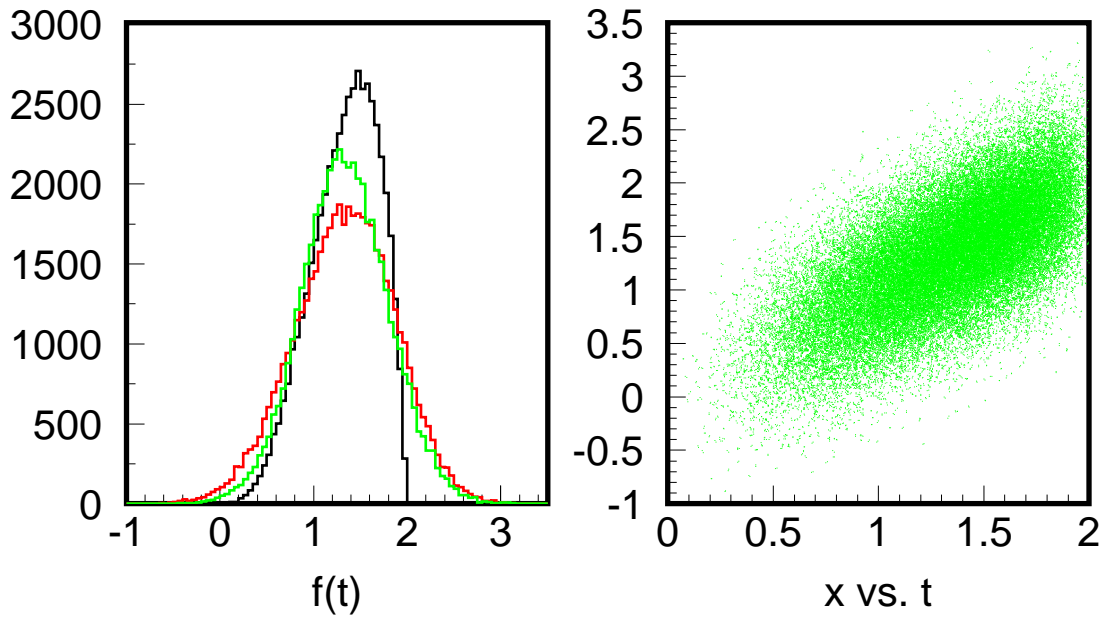}}
\caption[]
{Left: True distribution $f(t)$ (black) along with distribution of two 
measured quantities (green, red). Right: Correlation of one of the 
measured quantities with the truth.}
\label{example1}
\end{figure}

\subsection{\sf Fast function evaluation}
To speed up the learning process it is important that function evaluation
is as fast as possible in the inner learning loop. It can for example be
accelerated by tabulating the sigmoid function and using linear interpolation. 

\subsection{\sf Dynamic step size adaptation}
We use the following procedures to speed up learning:
From the first few thousand events the largest eigenvalue of the
Hessian (i.e. second derivative) matrix of $\chi^2$ with respect to the
weights is calculated according to the recipe of \cite{LeCun,Moller},
exploiting the power method and Rayleigh quotients \cite{Power}.
This defines the largest learning rate. 

\subsection{\sf Randomising event order}
During learning the network usually sees the events always in the
same order. Since the weights are updated every 200 or so events, 
the state of the network at the end of a training period has a slight bias
towards the last events. 
We have found it useful to randomise the order of the events in each iteration.
This avoids the bias to always be the same. Checking the network performance
on a test sample after each iteration one can observe and estimate the size 
of the statistical fluctuations due to this effect.
We have found that - especially in difficult learning patterns -
the randomisation of the order improves convergence. The chance 
that the network always moves along the same path and gets stuck in a 
local minimum is minimised. 

\subsection{\sf Choosing input variables}
There are two important classes of input variables that should be
considered: those which are correlated with the true output give
information about the deviation of the mean and median from 
the unconditional distribution, and quality variables which determine the
width of the conditional probability distribution, e.g. estimated
measurement errors. Both types should be fed into the net.  

\subsection{\sf Preprocessing of input variables}
A completely automatic and robust preprocessing procedure
has been developed. First all input variable distributions are equalised 
just like the output variable described above (by sorting).
This is especially important because otherwise 
(probably wrong or unreliable) extreme outliers in one variable can 
completely saturate neurons and thus dominate the net output.
The equalised variables are scaled to lie between -1 and 1. 
Using the inverse of the integrated $\chi^2$ function, the flat 
distributions are then converted into Gaussian distributions, centered
at zero with standard deviation 1. This is important for two reasons:
a nonzero mean value implies a large eigenvalue of the Hessian in weight 
space that restricts the initial allowed learning rate \cite{efficient}.
And a width of one, together with the random weights preset as described
above, makes sure that also the inputs to the output layer are distributed
with mean zero and width one, thus providing optimal conditions for a
fast initial learning and avoiding neuron saturation.   

\subsection{\sf Decorrelation of input variables}
Now all continuous input variables have a Gaussian shape with mean zero and
width one. However, they still may be correlated. For network
training it is advantageous to decorrelate the input variables. To do
this, first the covariance matrix of the preprocessed input variables 
is calculated. Then it is diagonalised using iterative Jacobian 
rotations \cite{blobel}. The rotated input vectors are divided by the 
square root of the corresponding eigenvalue. 
This way the covariance matrix of the transformed variables is a unit matrix.

\subsection{\sf Linear optimisation towards one input node}
Once the inputs are prepared this way, the covariance ``ellipsoid'' is a
sphere, and any rotation applied to the input vectors leaves them 
orthogonal and normalised. Now the correlation coefficent of 
each of the eigenvectors to the true target value is calculated. 
Using a sequence of $N-1$ rotations, it is possible to rotate this 
unit matrix such that only one variable shows the complete linear 
correlation to the target, and all other correlations are
zero. The resulting correlation is the best one can achieve with a linear
method. It also has been proven useful to rotate the complete correlation to
the second moment of the target distribution to the second input
variable etc. Exploiting the non-linear neural network, more is possible.  

\subsection{\sf An alternative mapping of the input variables}
Another possibility to increase net learning speed is to perform another
variable transformation: 
The mean performance is plotted against the equalised input variables.
If one observes a monotonuous but non-linear correlation, it is useful to
fit this dependence and take the fitted function value as input instead of 
the original variable.
Although this non-linear function should also be learned by the net, 
this preprocessing may help in finding the global minimum or at least
a good local minimum. 

\subsection{\sf Why learn the cumulative distribution?}
This algorithm learns the cumulative distibution in each output node,
not the interesting distribution itself. One may ask why this is necessary:
one could instead directly train the network outputs to the true $t$ value
lying in a given interval, since neural networks are able to map every 
function. This is correct. However, experience shows that is more difficult
for a network to learn it, especially when the correlation is weak. Some
more subtleties occur: The mean output sum of such a (multinomial) 
multi-output network is $1$, but event-by-event fluctuations lead to 
deviations. Smoothness is thus more difficult to achieve. Also, knowledge
of the cumulative distribtion readily allows for calculating expectation
values in a statistically optimal way.

\section{\sf Monte Carlo tests}

\subsection{\sf A simple toy example: two measurements of the same variable}

\begin{figure}[!th]
\mbox{\epsfxsize16.0cm\epsffile{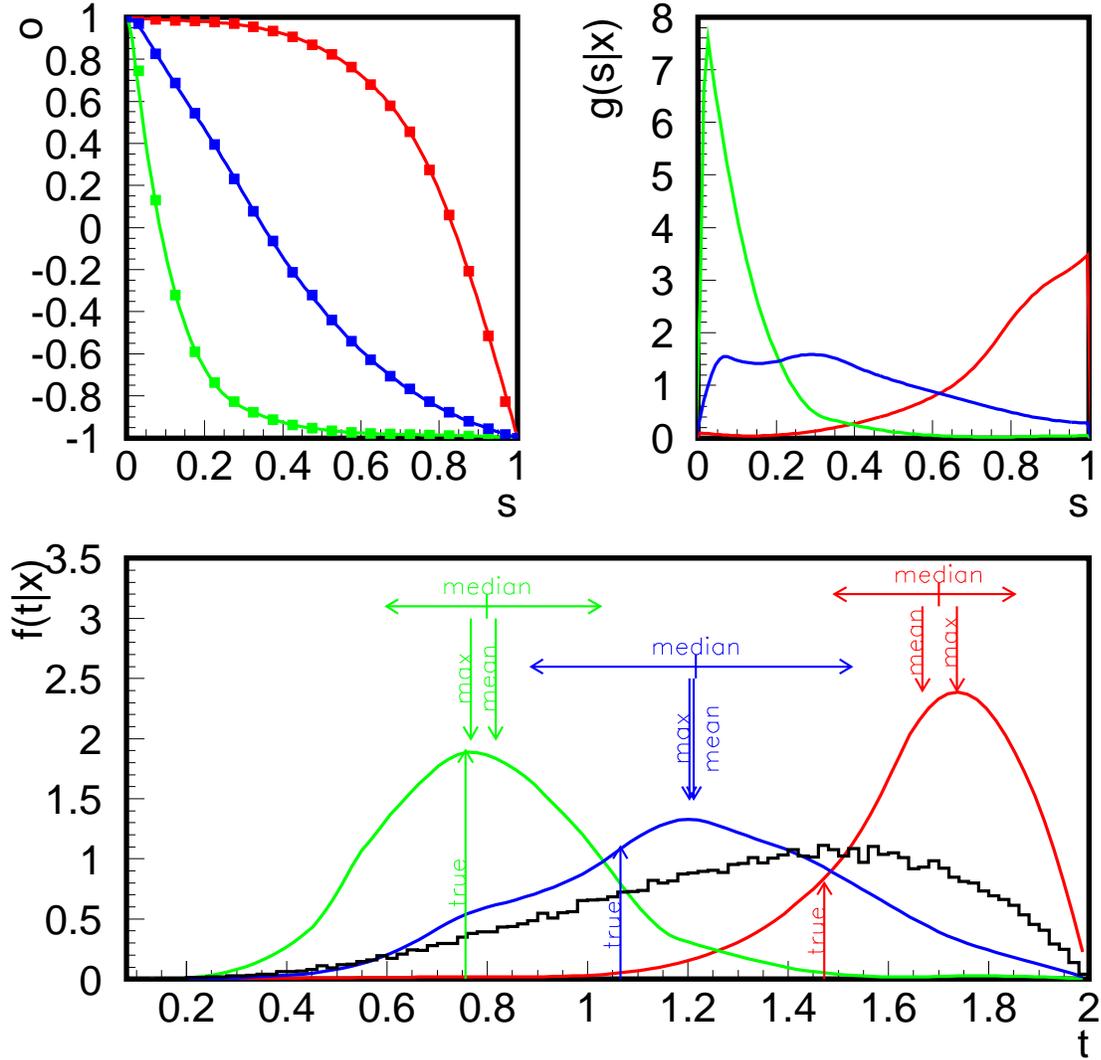}}
\caption[]
{Top left: Network output $o$ vs. purity $P(s)=s$ of output node for 
three events.
The boxes denote the 20 network outputs, the lines are the spline
fits $G(s|\vec{x})$. 
By construction, the unconditional $G(s)$ is the diagonal from $(0,1)$ to
$(1,-1)$. Top right: the conditional probability density 
$g(s|\vec{x})$ for the three events, determined from the first derivative 
of the
spline fit. Bottom: The conditional probability density $f(t|\vec{x})$ for the
three example events. In all cases also the true values, mean, median and 
left and right error are indicated.
The black histogram is the inclusive distribution $f(t)$.
}
\label{example2}
\end{figure}
\begin{figure}[!th]
\mbox{\epsfxsize16.0cm\epsffile{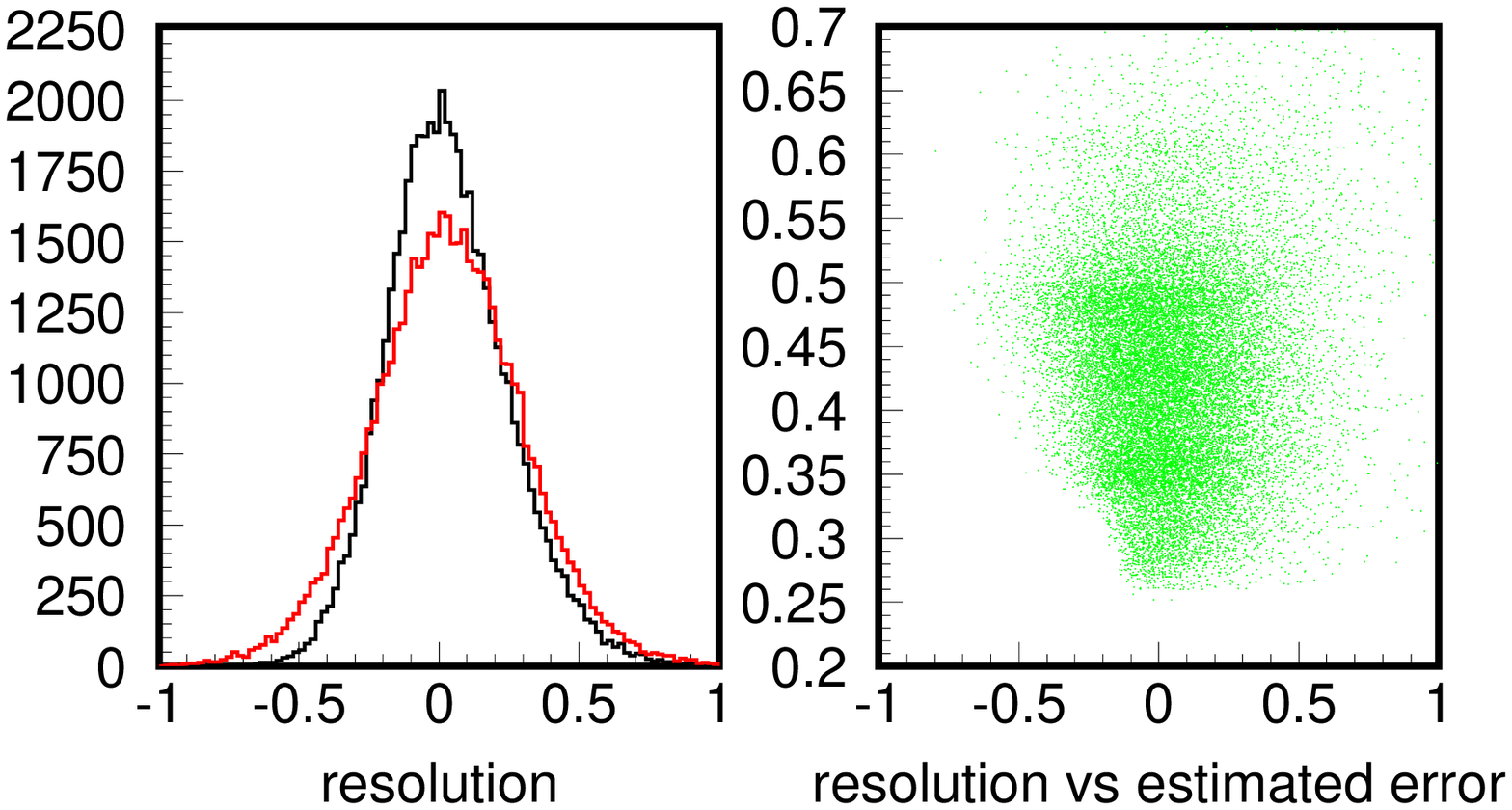}}
\caption[]
{Left: Resolution of the maximum likelihood estimate of $t$ (black), 
compared to the resolution of the weighted mean of both input 
measurements (red curve). 
Right: Error estimate ($\sigma_{left}+\sigma_{right}$) vs. deviation of
the maximum likelihood estimator from the true $t$ values.} 
\label{resolution}
\end{figure}
A Monte Carlo experiment shows how the method works and how it takes
into account the a priori knowledge of the inclusive distribution.
50000 events distributed according to $f(t)=5/8\cdot t^3\cdot(2-t)$ 
have been simulated. For each event two independent ``measurements'' 
with Gaussian smearing are available: One with a constant $\sigma=0.4$,
and another one with $\sigma=\sqrt{0.25^2+(0.35\cdot(2.-t))^2}$.
Fig.~\ref{example1} shows $f(t)$ along with the  distributions
of the two measurements. Observe that a part of the measurements    
yield values outside of the physical region. A simple weighted mean
cannot avoid this, but the Bayesian network will avoid it. 
The right side shows the correlation of the more precise measurement
with the truth. 
Four input variables are chosen, the two 
measurements and estimations (with $15\%$ smearing) of their uncertainties.
Of course, for the second measurement the error calculation cannot use the
true $t$, but has to estimate it from the very uncertain measurement.
The equalisation of the target variable is shown in Fig.~1.
\begin{figure}[!th]
\mbox{\epsfxsize16.0cm\epsffile{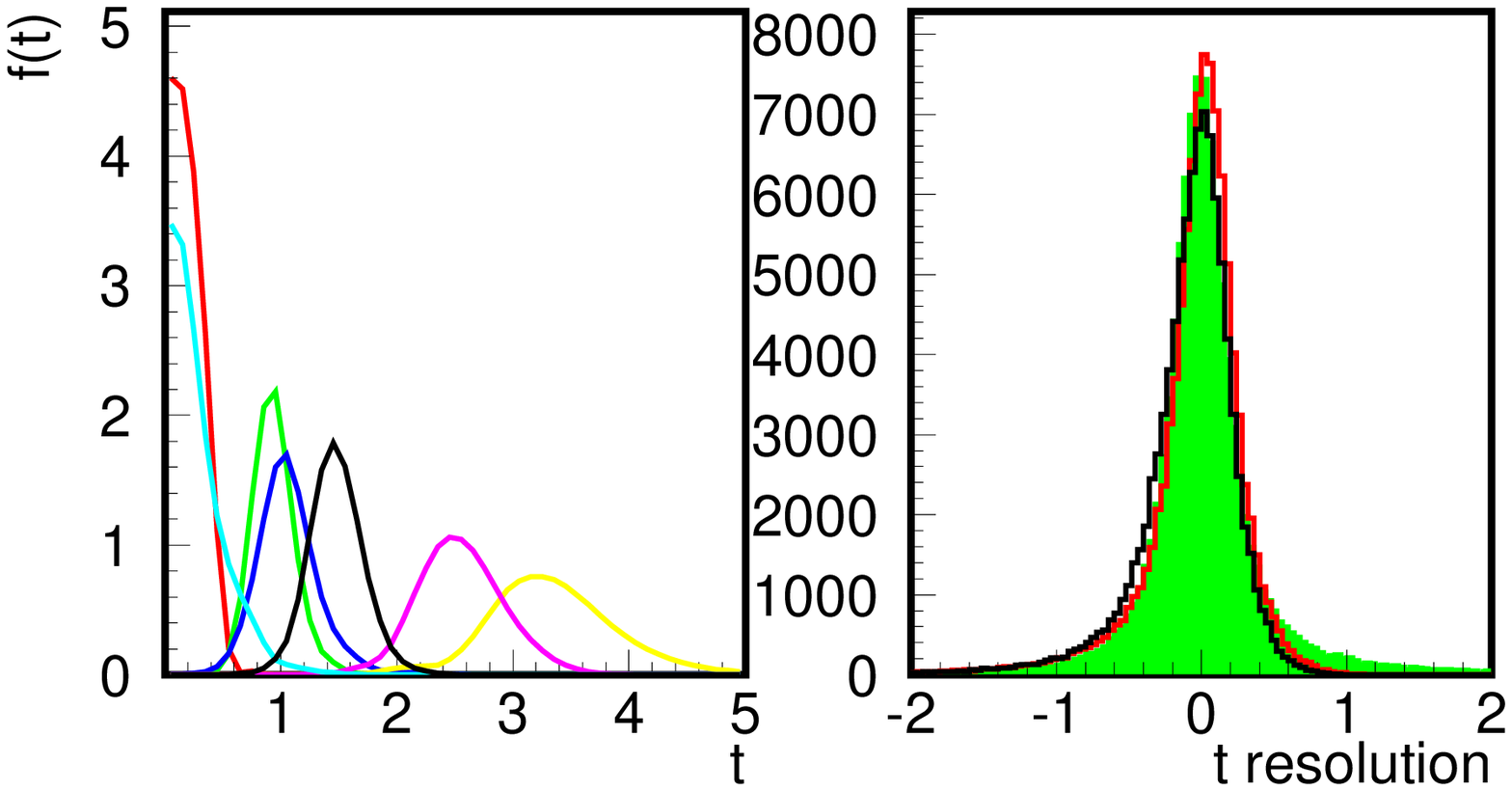}}
\caption[]
{NeuroBayes lifetime reconstruction from measured decay length and momentum.
\newline
Left: $f(t|\vec{x})$ for some events. The widths show the generally expected
tendency of getting broader with increasing $t$. Different widths at the
same mean $t$ are learned from the different combinations of $d$ and $p$
leading to the same $t$.   
\newline
Right:
Deviation of NeuroBayes median (black) and max likelihood (red) estimators from the true
$t$ values, compared to the direct ratio of measured variables (shaded). 
The resolution is similar, but the right tail vanishes due to the Bayesian
ansatz.
}
\label{lifetime}
\end{figure}
The output of three example events are shown in Fig.~\ref{example2}.
Two of the examples can well be reconstructed from the measurements
(steep curves top left), 
whereas the other only had weak additional information from
the measurements, such that the result differs only slightly from the
inclusive distribution.
Top right of the same figure shows the conditional probability densities
$g(s,\vec{x})$ of the transformed variable. A constant distribution here
means that there is no more information than the inclusive distribution.
One of the example events exhibits such a behaviour, only very low
and higher values are suppressed. On the bottom of Fig.~\ref{example2}
the final output $f(t|\vec{x})$ is shown for these example events, along
with their mean values, error estimates, median
and maximum likelihood estimate. Note that not only an expected value, but 
also an error estimate is given event by event. Deviations from Gaussian
behaviour is clearly visible, partly due to the limits of the physical
region.

Fig.~\ref{resolution} shows the resolution 
(distribution of estimated minus true values) of the maximum likelihood
estimator derived from $f(t|\vec{x})$, as compared to the weighted
average of the two measurements that enter the neural net. A marked
improvement is visible. Also the estimated uncertainty is important
and good information: Note how the resolution improves with a smaller 
error estimate (here taken as sum of left and right standard deviations).

\subsection{\sf More sophisticated: A function of two different measured 
variables with error propagation: lifetime reconstruction}
The proper lifetime $t$ of a relativistic particle can be reconstructed 
from the decay length $d$ and the momentum $p$ via
\begin{equation}
t=\frac{m}{c} \frac{d}{p}
\end{equation}
where $m$ and $c$ are constant and known quantities, the particle mass and 
the speed of light. The inclusive distribution is an exponential decay law:
$f(t)=e^{-\frac{t}{\tau}}$, where $\tau$ is the mean lifetime.  
The numbers taken for this example are typical for the measurement of
$b$-hadron lifetimes using inclusive reconstruction in DELPHI \cite{BSAURUS}.
The usual way is to measure $d$ and $p$ separately, and then to divide them.
However, both measurements can have relatively large errors and unphysical
values, in which case this is not the optimal thing to do.   

Here we investigate whether the net also can learn that it has to divide 
two input numbers and how to combine the measurement errors. Since three layer
neural networks can approximate any function this should also work in this
configuration.

A toy Monte Carlo has shown that this is feasable.
The momentum distribution is generated as in the toy example above, scaled
to between $0$ and $45\, GeV$, 
and decay length from an exponential distribution
in lifetime $t$ with $\tau=1.6\,ps$ and calculation of $d$.    
Decay length resolution is simulated as Gaussian smearing of width 
$250\, \mu m$, and momentum resolution as Gaussian smearing of
width $(10-p/5)\, GeV$.
The acceptance is modelled as  $A(d)= 1-e^{-\frac{d}{d_0}}$, in addition
events with negative measured decay lengths are rejected.
 
A neural network is then trained with just two input nodes, the measured
$d$ and measured $p$, and the true proper time $t$ to define the targets.  
The results are very promising: 
Some example events are shown in Fig.~\ref{lifetime}.
Obviously the net can learn the division and the error propagation. The
right plot shows the deviation of the NeuroBayes maximum likelihood and
median estimators along with that of the direct ratio of measured 
$d$ and $p$ (shaded area). They have about the same resolution, but do not
show the large upwards tail. 

This demonstrates that at least simple functional combinations 
of different measured quantities can directly be learned by such an
approach. This can be extremely helpful since also the (non-Gaussian) 
error propagation is handled correctly.  

\section{\sf Real application examples}
\subsection{\sf A difficult case: Improved B energy reconstruction in DELPHI}
b-hadrons may decay into thousands of different channels. BSAURUS 
\cite{BSAURUS} contains
algorithms that try to estimate the energy of a b-hadron inclusively, without
knowing which the actual decay channel was. The currently best estimate
\begin{figure}[!th]
\mbox{\epsfxsize16.0cm\epsffile{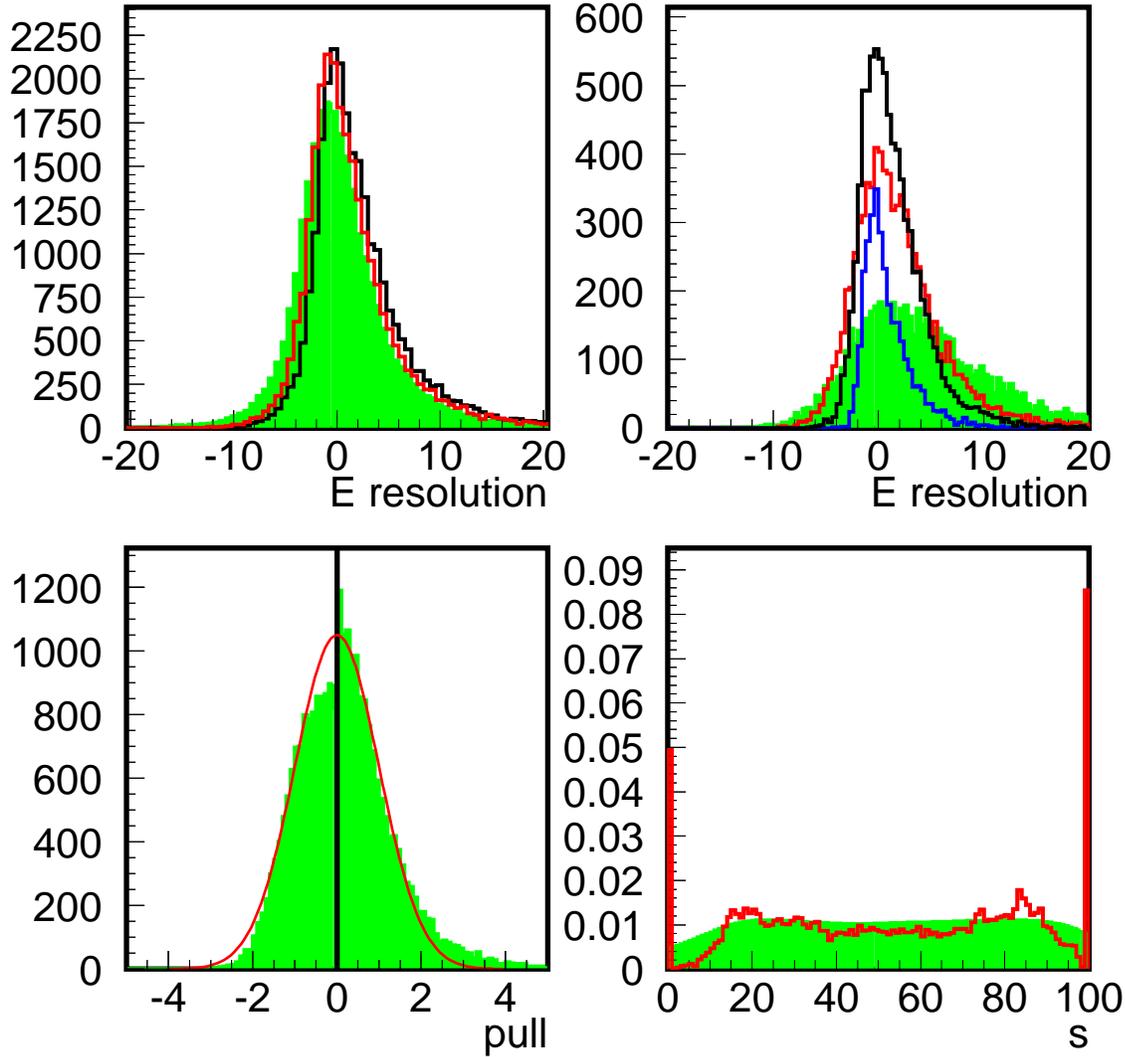}}
\caption[]
{NeuroBayes inclusive B energy reconstruction for the DELPHI experiment.
Top left: resolution of the currently best BSAURUS energy estimator
(shaded) with NeuroBayes median and max likelihood estimator. Top right:
resolution of NeuroBayes median estimator for 4 intervalls of its error
estimate: $\sigma<2.25\,GeV$, $2.25<\sigma<3\,GeV$, $3<\sigma<4\,GeV$ and
$\sigma>4\,GeV$ (shaded).
Bottom left: Pull distribution of the NeuroBayes median estimator.
$\sigma_L$ has been used when the estimated value was below the
true value and vice versa. It appoximates well the standard Gaussian of width
$1$.
Bottom right: Distribution of the maximal bins in $g(s|\vec{x})$ for many
events (dark histogram) 
and the mean $g(s|\vec{x})$ of all events (shaded area-
should be flat).    
}
\label{eb}
\end{figure}
is built from track rapidity, exploits the track neural network, the missing
energy in the hemisphere and the invariant mass of the particles which
probably stem from a B decay. A NeuroBayes network has been set up and trained
to try to improve its performance. Twenty input variables were chosen, which
include different estimators of the energy available in BSAURUS, their
input quantities as well as some measures of the measurement uncertainty    
like hemisphere multiplicity, reconstruction quality flag, B-fragmentation 
distinction quality, number of identified leptons and number of 
reconstructed jets.

Compared to the best classical BSAURUS estimator a non-negligiable 
improvement could be achieved (top left in Fig.~\ref{eb}.). 
Moreover, for the first time good error estimates are available:
This can be seen in the top right plot which shows the resolution in 
four bins of the estimated uncertainty. The corresponding resolutions
clearly are very different, it thus is possible to select events with
especially good reconstructed energy. This is very helpful, for example
in spectroscopy analyses.  
The lower left plot shows the pull distribution, i.e. the deviation of the
median energy estimate from the true value, divided by its estimated error,
$\sigma_l$ or $\sigma_r$, depending on the sign of the deviation. The
mean width is compatible with $1$ on both sides. However, the right side
shows a stronger, non-Gaussian tail which has its origin in low energetic
events. The single $f(t|\vec{x})$ distributions clearly exhibit this 
non-Gaussian behaviour.
The lower right plot shows the distribution of the maxima of $g(s|\vec{x})$,
which should and in the best of all worlds would be flat. It is however
clearly visible that the method often pushes the maxima into the extreme 
values $0$ and $1$, when the truth is near those values. The shaded area
shows the average $g(s|\vec{x})$, obtained by adding many events. This also
should be flat, and it is already much flatter than the maximum positions.   

\subsection{\sf A very difficult case: Econometry}
This algorithm is also very useful in time series predictions
\cite{honerkamp,masters}.  
An important application of time series prediction is financial 
forecasting, and maybe here the algorithm
brings the largest benefits since the 
correlations to the target are very small und hidden under large
stochastic background.
Historical data can be used to train the network. 

To predict the future of an equity price is of course a very 
difficult problem,
since it is very uncertain and clearly depends on many unforeseeable
facts. The most robust estimate is that the mean value will stay constant.
On the other hand, it is known that over many years there is a rise 
observable. A first look to the data suggests that the ``random walk model''
works very well. This model suggests that every day's price change is 
truely random with mean value zero, e.g. according to a Gaussian distribution 
\begin{equation} 
f(t,\vec{x})=f(t)=\frac{1}{2\pi\sqrt{\sigma}} e^{-\frac{t^2}{2\sigma^2}}
\end{equation}   
and there is no more information in the price history from which
$\vec{x}$ can be constructed.
This states that it is impossible to predict the direction, 
but the standard deviation $\sigma$ (the ``volatility'') can 
be estimated from history. Lots of physicists and mathematicians are
now hired by banks to do quantitative risk calculations and develop
strategies \cite{bouleau}.  
 
Nevertheless some features which probably lie in human 
nature (greed and fear) make it possible to look a bit into the future.
This is the basis of what is known as ``technical analysis''.
There is an enormous literature on this subject, see e.g.
\cite{murphy}. Many of the common ``technical'' concepts seem a bit 
esoteric and subjective to a physicist and there is very few 
quantitative information about the reliability published.
 
Without going into details -- just saying
that a clever choice of input variables with intelligent preprocessing 
is an important prerequisite -- it is possible to observe short time 
correlations from the past to the future of equity prices.
Of course it must be made sure that enough input data is taken, 
such that the net does not only learn statistical fluctuations "by heart".

\begin{figure}[!th]
\mbox{\epsfxsize16.0cm\epsffile{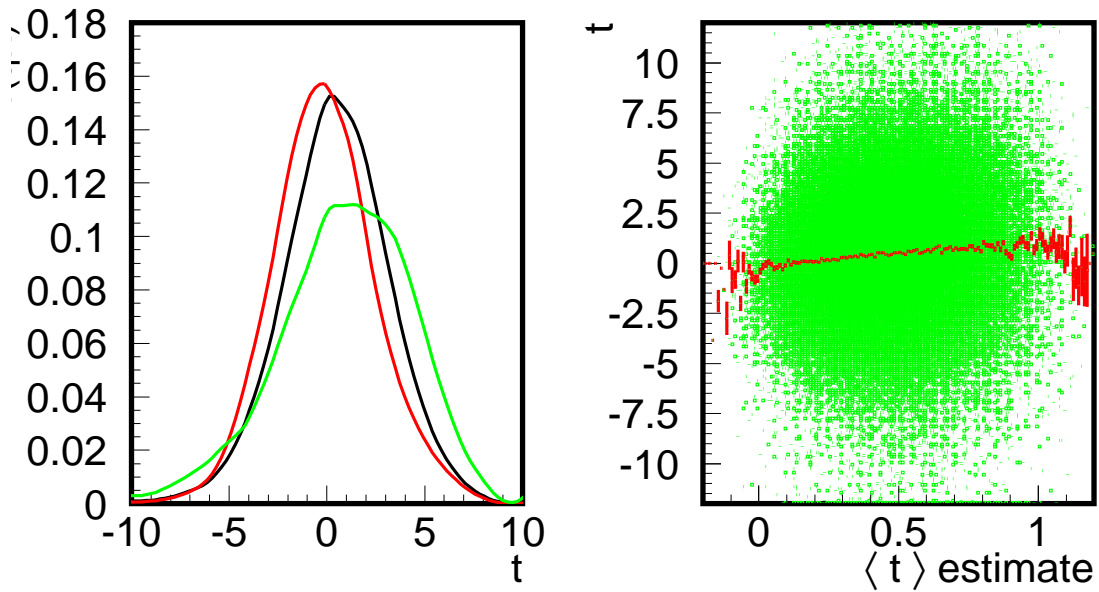}}
\caption[]
{Left: Three examples of estimated conditional probabilities for 
a measure for 10 working day price changes of American Dow Jones equities. 
Right: Correlation of the true 10 days' performance to the estimated 
mean performance. A clear correlation is visible.  
}
\label{finperformance}
\end{figure}

Training on 20 years' data of the 30 equities of the Dow Jones Industrial Index
with a relatively simple input model leads to the performance shown in 
Fig.~\ref{finperformance}. 
A clear dependence of the mean true performance as function of the mean of
the estimated performance is visible, which still is small, but not at all
negligible compared to the spread. The left side of that figure shows the
resulting conditional probability densities for three examples.
In addition to finding optimal combinations of different technical
indicators on the basis of history,  
NeuroBayes does not only give an indication of ``up'' or ``down'' but
quantifies it day by day and equity by equity. This could help in 
decision finding and especially timing of financial transactions.  

The method may also be used for determining a fair price for a call option
when $t$ is the price of the underlying at the date of maturity,
the distribution of which is estimated by $f(t)$, 
and $S$ is the strike price: Neglecting interest rates (can be included),
$a(t)= min(t-S,0.)$. The famous Black-Scholes formula
\cite{blackscholes} is just a special
case of this pricing model for the random walk model shown above,
i.e. a simple Gaussian with zero mean $\langle t\rangle = 0$ and 
``volatility'' $\sigma$.

\section{\sf Developments from 2001 to 2003}
The NeuroBayes algorithm has been completely recoded in a structured
way by the Phi-T project, J. Bossert, A. Heiss and the author, between 
2001 and 2002. Phi-T was sponsered by the exist-seed program of the 
German Bundesminsterium f\"ur Bildung und Forschung BMBF with the aim
to found a company which makes the technology available outside
physics.
In October 2002, the Phi-T Physics Information Technologies GmbH was
founded in Karlsruhe.
Phi-T has developed lots of additions and improvements to the
algorithm and the code, bindings for several programming languages and
several input/output interfaces.
They use NeuroBayes with large success for insurance and
banking applications \cite{Phi-T}.      

Also in physics research NeuroBayes and its predecessors have found a
lot of very successful applications:
\begin{itemize}
\item Measurement of the b-fragmentation function with DELPHI
  \cite{ramler, fragmentation}
\item Precision lifetime measurements of $B^+$ and $B^0$ mesons \cite{blifetime,haag}
\item Spectroscopy of orbitally excited B mesons, in particular 
    discovery of $B_S^{**}$ mesons \cite{B**,albrecht}
\item Search for resonances in the final state $B\pi^+\pi^-$ with
  DELPHI \cite{huegel},
\item Electron identification in the CDF experiment at Fermilab
  \cite{milnik}

\end{itemize}

\section{\sf Summary}
A simple but effective and robust algorithm has been described that
calculates the conditional probability density $f(t|\vec{x})$ from
(simulated or historical) example events for multidimensional
input data. It is using a smoothing function over a number
of neural network outputs, each of them delivering a Bayesian
a posteriori probability that the real value lies above the corresponding
threshold. Completely automatic and parameter-free input variable 
preprocessing and stochastic second order methods to adapt
individual learning rates for each weight speed up the quasi-online 
learning by at least an order of magnitude compared to plain-vanilla
backpropagation. The generalisation ability is controlled by a weight
decay regularisation with several independent regularisation constants
that all are chosen and recalculated online during the training using
a Bayesian reasoning. 

Several toy examples and some real examples have been tested. 
The procedure is an event-by-event unfolding with multidimensional input.
It can directly be used in maximum likelihood fits. It allows to handle 
difficult measurements in a large number of application 
areas in an optimal way.

\section*{\sf Acknowledgements}
I like to thank V. Blobel (Univ. Hamburg) for many good algorithms 
and enlightening discussions, Z. Albrecht and U. Kerzel
(Univ. Karlsruhe) for their contributions,
G. Barker and M. Moch (Univ. Karlsruhe) for carefully reading the 
mansucript and T. Allmendinger, C. Weiser (Univ. Karlsruhe) and 
N. Kjaer (CERN) for useful feedback.
Last but not least I thank the Phi-T team J. Bossert, C. Haag, 
A. Heiss and L. Ramler for making NeuroBayes a success. 


\end{document}